\documentclass[aps,prl,twocolumn,showpacs]{revtex4}
\usepackage{graphicx}

\begin{document}


\title{Strong spin-orbit splitting on Bi surfaces}
\author{Yu. M. Koroteev$^{1,2}$, G. Bihlmayer$^3$, J. E. Gayone$^{4,5}$,
E. V. Chulkov$^{1,6}$, S. Bl{\"u}gel$^3$, P. M. Echenique$^{1,6}$, and
Ph. Hofmann$^{4,*}$}
\affiliation{$^1$Donostia International Physics Center (DIPC),
 20018 San Sebasti{\'a}n/Donostia, Basque Country, Spain\\
$^2$Institute of Strength Physics and Materials  Science, Russian
Academy of Sciences, 634021, Tomsk, Russia\\
$^3$Institut f\"ur Festk\"orperforschung, Forschungszentrum
J\"ulich, D-52425 J\"ulich, Germany\\
$^4$Institute for Storage Ring Facilities, University of Aarhus,
DK-8000 Aarhus C, Denmark\\
$^5$Centro Atomico Bariloche and CONICET, 8400 S.C. de Bariloche, Argentina\\
$^6$Departamento de F\'{\i}sica de Materiales and Centro Mixto CSIC-UPV/EHU,
Facultad de Ciencias Qu\'{\i}micas, UPV/EHU, Apdo.~1072, 20080
San Sebasti{\'a}n, Basque Country, Spain}
\email[]{philip@phys.au.dk}\homepage[]{http://www.phys.au.dk/~philip/}
\date{\today}

\begin{abstract}
Using first-principles calculations and angle-resolved photoemission,
we show that the spin-orbit interaction leads to a strong splitting of
the surface state bands on low-index surfaces of Bi. The
dispersion of the states and the corresponding Fermi surfaces are
profoundly modified in the whole surface Brillouin zone.
We discuss the implications of these
findings with respect to a proposed surface charge density wave on
Bi(111) as well as to the surface screening, surface spin-density waves,
electron (hole) dynamics in surface states, and to possible applications
to the spintronics.
\end{abstract}


\maketitle

Recently, spin-orbit coupling (SOC) on surfaces and the resulting
splitting of the surface-state bands has attracted considerable
attention. While it is a well-established fact that the reduction of
coordination at surfaces and in thin films can lead to pronounced magnetic
effects, the discovery of a small splitting in the band of
the $sp$ surface state on the non-magnetic Au(111) surface and its
interpretation as being due to SOC by LaShell {\em et al.}~\cite{LaShell96}
came as a surprise. More sophisticated angle-resolved photoemission
(ARPES) investigations and calculations have meanwhile confirmed this
splitting \cite{Reinert02} and the combination of the experimental results
with first-principles calculations do indeed proof that the SOC is
causing it \cite{Reinert02,Henk03}. Later, larger SOC-induced
splittings were found on the Li-covered surfaces of W and Mo
\cite{Rotenberg99} and the predicted difference in spin-orientations for H on W(110)
was confirmed experimentally using spin-resolved ARPES \cite{Hochstrasser02}.
Since these surface states contribute only very little to the density of
states at the Fermi level, the observed spin-orbit or Rashba splitting of these states
will not show up in transport phenomena. On the other hand, surface states of a
semimetal would give a prominent contribution \cite{Hengsberger00,Agergaard01} which
could make these systems interesting for applications in the field of spintronics.
The surfaces of the semimetal Bi seem to be ideal to advance our understanding of
SOC on surfaces and how it manifests itself in experiments. Of
particular interest are the influence of the SOC on the
electron-phonon coupling \cite{Rotenberg00}, electron and hole dynamics
\cite{Echenique04}, and the possible formation of surface charge (spin)
density waves. The occurrence of strong SOC in low-dimensional structures
of non-magnetic materials could also have applications like spin-filter devices.

ARPES measurements of the Fermi surface (FS) and surface states were recently
performed by Ast and H{\"{o}}chst for Bi(111) \cite{Ast01}. They interpreted
the obtained FS in terms of two different surface bands
which are not degenerate at the $\bar{\Gamma}$ point. Based on this
electronic structure they proposed a possible mechanism for the formation
of surface charge density waves (CDW) on Bi(111) \cite{Ast03}.
Agergaard {\em et al.} \cite{Agergaard01} measured surface states and
the FS on Bi(110). They pointed out that these surface
states should be completely non-degenerate because of spin-orbit
splitting but, as we show below, the splitting is so large that
an easy identification of the spin-split bands was not possible.

Bulk Bi is a semimetal where the strong spin-orbit interaction essentialy
accounts for the existence of the hole FS at the T point \cite{Gonze90}
but does not lead to any lifting of the spin degeneracy because of
inversion symmetry. However, the high atomic number of Bi (83) and
the pronounced splitting in the atomic $p$ levels (the atomic $p_{3/2}$-$p_{1/2}$
splitting in Bi (1.5 eV) is three times stronger than in Au (0.47 eV)
\cite{Moore49}) should lead to an observable splitting of the surface
state bands.

In this Letter we show that indeed the surface states on low-index
surfaces of Bi exhibit a spin-orbit splitting of the bands which is
by far stronger than any case reported so far.
We prove this by combining the results of first-principles
calculations with high-resolution measurements of the electronic
structure by ARPES. The results of the calculation agree well with
experiment but only if the SOC and hence the removal of the spin
degeneracy is taken into account. We find that the SOC induced
splitting is an essential ingredient for the description of the
electronic structure: it profoundly changes the surface-state
dispersion and the corresponding Fermi surfaces on all the Bi
surfaces of interest. In particular, it is responsible for the existence
of the six FS hole lobes in the $\bar{\Gamma}$$\bar{\mathrm M}$ symmetry directions
for Bi(111), and it also leads to correct dimensions
of the electron FS hexagon around the $\bar{\Gamma}$ point.
Our relativistic calculation demonstrates that despite the existence
of nesting at the electron FS hexagon of Bi(111)  the formation of a CDW
\cite{Ast03} appears to be improbable since this nesting couples
the states with the same energy but different spin,
$\epsilon(\mathbf{k},\uparrow)=\epsilon(-\mathbf{k},\downarrow)$. It could
lead rather to the formation of a spin-density wave and not to a CDW.

Here we present results for three surfaces:
Bi(111), Bi(110) and Bi(100). The calculations have been performed
by using the full-potential linearized augmented planewave method in
film-geometry~\cite{Wimmer81} as implemented in the {\sc fleur} program
and local density approximation for the description of
exchange-correlation potential. SOC is included
self-consistently as described in Ref.~\cite{Li90}.
All the Bi surfaces were simulated by a 22 layer film  embedded in
vacuum. One side
of the film was terminated with hydrogen to avoid interaction between
the surface states of the two surfaces of the film. The H atoms were
placed in a distance of $\approx$ 2~\AA~ from the Bi. On the other side
of the film, the termination was chosen such that the interlayer distance
between surface and subsurface atoms was the shorter one of two
possible terminations. For the calculations
a planewave cutoff of $K_{\rm max}=3.4$~(a.u.)$^{-1}$ was used and
the surface Brillouin zone (SBZ) was sampled with up to
121 {\bf k}$_{\|}$-points.

We also show surface states measured with angle-resolved photoemission
spectroscopy. The experiments were performed at the SGM-3 beamline of
the synchrotron radiation source ASTRID in Aarhus \cite{Hoffmann03}.
All surfaces were prepared from mechanically polished
single-crystal surfaces which were cleaned in situ by cycles of Ne ion
bombardment and annealing to about 473 K. This resulted in well-ordered
and clean surfaces as judged by low energy electron diffraction and
Auger electron spectroscopy, respectively.
The total energy resolution for the data
shown below is better than 35 meV. The angular resolution of the analyser
is $\pm0.5^{\circ}$. The samples were cooled to approximately 30 K.

Before discussing our results, we briefly explain the
symmetries relevant for the spin-orbit splitting of electronic
bands. Time reversal symmetry requires that
$\epsilon(\mathbf{k},\uparrow)=\epsilon(-\mathbf{k},\downarrow)$.
This means that if one has a surface state at $\mathbf{k}$ with a
binding energy $\epsilon$ and a spin $\uparrow$ then there must also be a
state at $-\mathbf{k}$ with the same energy but spin
$\downarrow$. This has to be combined with the usual space group
symmetry. The combination has two consequences. First, if the space
group contains inversion symmetry
($\epsilon(\mathbf{k},\uparrow)=\epsilon(-\mathbf{k},\uparrow)$) the bands
are obviously doubly degenerate. This can happen in the bulk but not
at the surface. Second, in the case of surface states,
the splitting has to be zero at some special points of the surface
Brillouin zone. The application of time-reversal symmetry alone yields
that one of these points is the centre of the zone $\bar{\Gamma}$
for which  $\mathbf{k}_{\parallel}=0$. This is indeed observed in the
dispersion of the spin-orbit split states on Au(111)
\cite{LaShell96}. The combination of time-reversal symmetry with
translational symmetry dictates that the spitting must also be zero
for any point
which is situated half-way between two $\bar{\Gamma}$ points.

\begin{figure}
\includegraphics[scale=0.8]{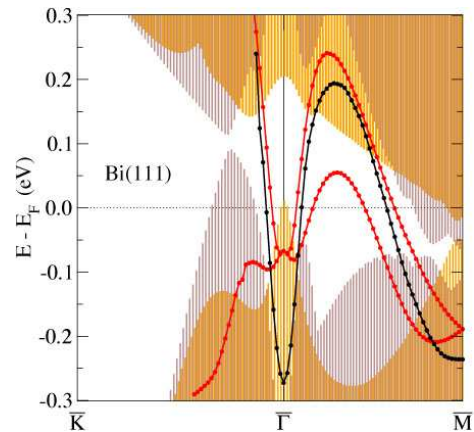}
  \caption{ Surface states of  Bi(111) calculated without
            (black) and with (red) spin-orbit splitting included.
            The shaded areas show the projection of the bulk bands
            obtained without (violet) and with (yellow) SOC and their
            superposition (brown).
  \label{fig:1}}
\end{figure}

Fig.\ref{fig:1} shows the electronic structure of Bi(111) together with
the projected bulk band structure for the (111) surface calculated with and
without SOC. The projection was calculated using the tight-binding
model of Ref. \cite{Liu95}. For clarity, we only show
the surface states which are located on the clean surface of the slab.
In the case without SOC, we find a parabolic $\bar{\Gamma}$
surface state located in
the nonrelativistic energy gap. Around $\bar{\Gamma}$ this surface state
band gives an electron FS hexagon. When the SOC
term is included it results in a spin-splitting of the surface state in all
the symmetry directions and leaves it degenerate only at $\bar{\Gamma}$
and at $\bar{\mathrm M}$. The latter is expected because $\bar{\mathrm{M}}$ is
a high-symmetry point on the SBZ boundary which lies, in contrast to e.g.\
$\bar{\mathrm K}$, in the middle between two $\bar{\Gamma}$ points.
Around $\bar{\Gamma}$ this relativistic surface
state is degenerate with bulk states and shows less clear surface character.
The lift of the spin degeneracy leads to radical change of the surface FS:
1) The radius of the FS hexagon is smaller by 30\% compared to the nonrelativistic
calculation; 2) In the $\bar{\Gamma}$$\bar{\mathrm M}$ symmetry directions
the hole lobes are formed.
Another remarkable feature of the Bi(111) surface electronic structure is the very
strong anisotropy of the spin-orbit splitting: it is $\approx $0.2 eV
in the $\bar{\Gamma}$$\bar{\mathrm M}$ direction and even more in the
$\bar{\Gamma}$$\bar{\mathrm K}$ one.

\begin{figure*}
\includegraphics[scale=0.85]{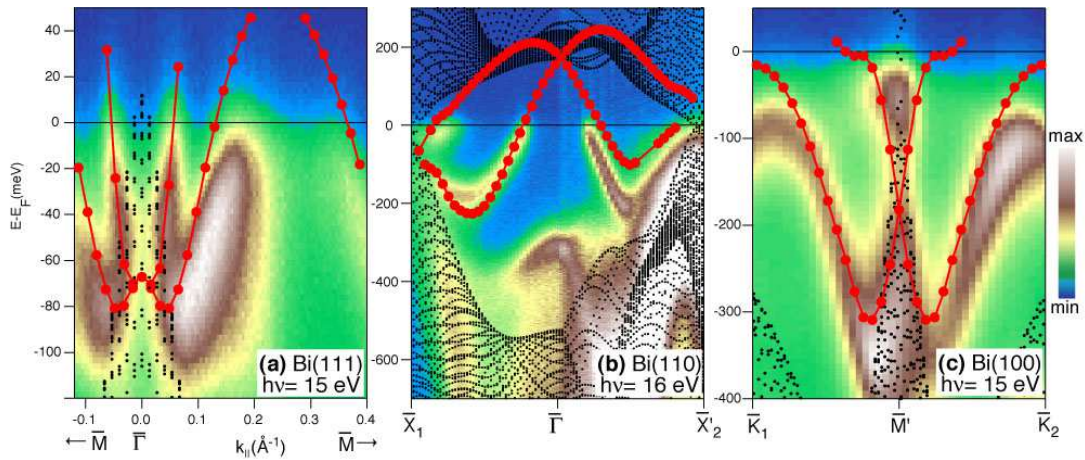}
  \caption{Calculated and measured electronic structure in the vicinity
  of two high  symmetry points on three surfaces of Bi. (a)
  $\bar{\Gamma}$ on Bi(111), (b) $\bar{\Gamma}$ on Bi(110), and (c)
  $\bar{M_1}$ on Bi(100). The small black dots are the projected bulk band
  structure calculated using the tight-binding model of Liu and Allen
  \cite{Liu95}. The red filled circles are the calculated surface state
  energies, thin red line is a guide to the eye. The photoemission intensity
  is linearly scaled from dark blue (minimum) to white (maximum).
  \label{fig:2}}
  \end{figure*}

The occurence of strong spin-orbit splitting is confirmed when the
calculations are compared to the experimental results. Fig. \ref{fig:2}
shows the calculated electronic structure for the three surfaces
together with experimental data. In Fig.~\ref{fig:2}(a) we compare
the results for Bi(111) and find excellent
agreement for the two split surface states near $\bar{\Gamma}$. The
experimental results also agree with recently published data for
Bi(111) \cite{Ast01} but the two split bands
appear better resolved here. Here, as in Ref. \cite{Ast01}, the
intensity of both surface states strongly decreases close to
$\bar{\Gamma}$. This is most likely due to the overlap with the
projected bulk band structure. The surface states are no longer genuine
surface states but surface resonances which penetrate much more deeply
into the crystal and give a lower photoemission intensity.
Although it is therefore not simple to decide if the states are in fact
degenerate at $\bar{\Gamma}$ or not, we find no evidence for the latter.
This discrepancy with the data of Refs. \cite{Ast01,Ast02}
is most likely due to a sample misalignment which
later was found to have been present \cite{Ast03}.
Our interpretation is in disagreement with the result
from Ast and H\"{o}chst who used a Bi(111) bilayer to simulate
the surface state dispersion of Bi(111) with a tight-binding model
\cite{Ast03b}. In their case, the calculated bands match the experimental
dispersion  beyond the $\bar{\Gamma}$ point when the bilayer thickness is
increased by 70\% with respect to the bulk value  and when the SOC
strength was reduced to 13\% of the experimental value 1.5 eV \cite{Moore49}.
The bilayer, however, always has inversion symmetry and therefore this calculation
yields two spin-degenerate bands near the Fermi energy that do not cross
at $\bar{\Gamma}$.

Fig. \ref{fig:2}(b) shows the situation near $\bar{\Gamma}$ of Bi(110).
The experimental results have already been published elsewhere \cite{Agergaard01}.
In the theory one can clearly see that, as in Bi(111), the surface state on
Bi(110) is degenerate at $\bar{\Gamma}$ and splits into two surface states
along the symmetry lines with one electron per $\mathbf k$-point.
In contrast to Bi(111) this surface state is unoccupied at $\bar{\Gamma}$ and
has negative effective electron masses that lead to the formation of the
hole FS pocket around $\bar{\Gamma}$ \cite{Agergaard01}. This specific behavior
of the surface state bands is also responsible for the formation of the electron
FS pocket  between $\bar{\mathrm{X}}_1$ and
$\bar{\mathrm{M}}_1$ and the hole pocket at $\bar{\mathrm{M}}_1$ \cite{Agergaard01}.
In the experiment only the lower branch of the spin-orbit split
state can be observed as it enters the occupied states. Such a
situation can be highly confusing because the band could be mistaken for a
simple parabolic hole pocket.

The scenario of a very steep band and a flatter one near
$\bar{\Gamma}$ can also be found on Bi(100). This is shown in Ref.
\cite{Gayone03} and is therefore not presented here. Instead,
Fig. \ref{fig:2}(c) shows the situation near the $\bar{\mathrm{M}}'$ point.
This point is the $\bar{\mathrm{M}}$ point of the quasi-hexagonal SBZ of
Bi(100) which is not lying on the mirror plane of the SBZ (see Ref.
\cite{Gayone03}). Since all $\bar{\mathrm{M}}$ points of a (quasi)
hexagonal SBZ fulfill the criterion of lying exactly in the middle of
the line joining two $\bar{\Gamma}$ points, we also expect a
degenerate surface state here. As Fig. \ref{fig:2}(c) shows, this is
indeed the case. In fact, here the bands close to the high symmetry
point are so steep that the dispersion can not be resolved in the
experiment. In the rest of the SBZ the agreement between experiment
and calculation is more difficult to find. This is due to the deep penetration of
the surface states into the bulk, such that even a calculation with a
22 layer film cannot completely avoid the interaction between the two surfaces.
Details of the electronic structure of Bi(100) will be published
elsewhere \cite{Gayoneunpubl}.

The spin-orbit splitting obtained for the Bi surfaces is a few times bigger
than that of the surface state on Au(111) \cite{LaShell96,Reinert02}, which
is to be expected since the atomic
spin-orbit splitting in Bi is three times larger than that in Au.
However, the character of the spin-orbit splitting in the present case
is different from that in Au(111). For the latter, the electronic
structure can be described by a free electron-like two-dimensional
surface state at $\bar{\Gamma}$. In such a case the spin-orbit interaction
can be treated by adding a so-called Rashba term to the non-relativistic
Hamiltonian \cite{Petersen00}. This leads to a splitting of the surface
state which is linear in $\mathbf k$.
In the case of Bi the surface states are not free electron-like and they
are distributed over the whole SBZ, therefore the spin-orbit splitting
shows a much more complex behavior which can only be revealed
by first-principle band structure methods.

The SOC-induced  splitting should have some important consequences
for the physical properties of the Bi surfaces, in particular,
for the screening. In the Lindhard picture of screening,
the susceptibility $\chi(\mathbf{q})$ is essentially given by an integral
over all processes where an electron hops between an occupied state
and an unoccupied state separated by $\mathbf{q}$. In a two-dimensional
situation this type of screening can lead to a CDW-type instability
only when there are `nested' elements of the FS, separated by $\mathbf{q}$.
Such a situation exists for Bi(111) where the FS of the electron pocket
around $\bar{\Gamma}$ is hexagonal \cite{Ast03}. Ast and H\"{o}chst have
recently shown \cite{Ast03} that the leading edge of the energy
distribution curves at the Fermi level crossing shifts
discontinuously as a function of temperature, indicating the formation
of a CDW. However, when we take into account the spin of the states involved
in the alleged formation of the CDW, the electron hopping across the
FS would have to undergo a spin-flip because of the split nature of the bands.
This makes the occurrence of a CDW very unlikely.

The spin-orbit splitting in surface bands on the Bi surfaces can also have
drastic consequences for electron and hole dynamics in surface states.
In particular, the surface response function should include all
the spin-flip processes between the split surface bands with different spin
direction. It can lead to the formation of surface spin-density waves even
in cases when the nesting at the surface FS does not occur. The spin-orbit
splitting should also lead to different hole (electron) lifetimes in surface
states compared to that for the non-split surface state. This is due to both
the surface response function that now includes spin-flip processes and
to a different phase space factor \cite{Echenique04,Kliewer00}.
The surface state spin-orbit splitting can also affect the electron-phonon ($e$-$ph$)
coupling on the Bi surfaces. Strong $e$-$ph$ coupling in surface states on Bi(100)
was already discussed by Gayone {\em et al.} \cite{Gayone03} who assumed direct
interaction between electron and phonon systems. Here we would like to note that
the existence of spin split surface states also permits a spin-wave
mediated $e$-$ph$ interaction.

The lack of inversion symmetry and the large spin-orbit splitting holds also
at Bi-insulator interfaces where spin-orbit split surface states turn
to interface states. We speculate that ultra-thin films of Bi covered with an
insulator become much more effective spin-filters and spin-manipulators for injected
spin-polarized electrons than semiconductors explored in spintronics because
the relevant $k$-vectors and the difference of the $k$-vectors for spin-up and
spin-down electrons are much larger than for semiconductors.

In conclusion we have presented the first {\em ab initio} calculation of the relativistic
surface electronic structure for Bi(111), Bi(110), and Bi(100). We have shown that
the SOC term leads to a strong and anisotropic splitting of the surface state
bands that profoundly modifies the dispersion of the surface states and
the surface FS. The calculated results are in good agreement with
the experimental data if the spin-orbit splitting is taken into account.
We have discussed the possible effect of the SOC interaction on the
surface response function and new mechanisms of electron and hole decay in
the Bi surface states in terms of electron-electron and electron-phonon
interactions. We also discussed the possible use of this effect for
spintronic applications.

\begin{acknowledgments}
This work was partially supported by the UPV/EHU, Spanish MCyT, the MAx Planck
Research Award Funds, and by the Danish National Science Foundation.
\end{acknowledgments}


\end{document}